\documentclass[aps,prb,showkeys,showpacs]{revtex4}

\usepackage{graphicx}
\usepackage{amsmath}
\def\lsim{\mathrel{\rlap{\lower5pt\hbox{\hskip0pt$\sim$}}
\raise1pt\hbox{$<$}}}                

\begin{document}

\title[Elastic contact between self-affine surfaces]{Elastic contact between self-affine surfaces: Comparison of numerical stress and contact correlation functions with analytic predictions
}

\author{Carlos Campa\~n\'a,$^{1,2}$ Martin H. M\"user$^2$ and Mark O. Robbins$^3$}
\address{$^1$CANMET-Materials Technology Laboratory, Natural Resources Canada,
Ottawa, Ontario, K1A 0G1, Canada}
\address{$^2$Department of Applied Mathematics,
University of Western Ontario, London, ON, Canada N6A 5B7}
\address{$^3$Department of Physics and Astronomy,
Johns Hopkins University, Baltimore, MD 21218} 

\date{\today}
\begin{abstract}
Contact between an elastic manifold and a rigid
substrate with a self-affine fractal surface is
reinvestigated with Green's function molecular dynamics.
The Fourier transforms of the stress and contact autocorrelation functions
(ACFs) are found to decrease as $|{\bf q}|^{-\mu}$ where ${\bf q}$ is the
wavevector.
Upper and lower bounds on the ratio of the two correlation functions are used to
argue that they have the same scaling exponent $\mu$.
Analysis of numerical results gives $\mu =1+H$,
where $H$ is the Hurst roughness exponent.
This is consistent with Persson's contact mechanics theory, while asperity models give $\mu= 2(1+H)$.
The effect of increasing the range of interactions from a hard sphere repulsion to exponential decay is analyzed.
Results for exponential interactions are accurately described by recent
systematic corrections to Persson's theory.
The relation between the area of simply connected contact patches and the normal force is also studied.
Below a threshold size the contact area and force are consistent with Hertzian contact mechanics, while area and force are linearly related in larger contact patches.
\end{abstract}


\maketitle

\section{Introduction}

The distribution of pressures in the contact between two solids
has a crucial influence on the amount of friction and wear that 
occur when the solids slide against each other.
As a consequence, there is great interest in reliable predictions for the
dependence of the distribution on the externally imposed load $L$, the mechanical
properties of the solids in contact, and their surface 
topographies~\cite{bowden56}.
Of course, it would be desirable to have  theories at hand that
allow one to calculate these distributions analytically or 
numerically at a moderate amount of computing effort.
While solving numerically the full elastic or plasto-elastic behavior
of contacting solids has become an alternative to studying analytical theories,
{\it numerically exact approaches} incorporating 
long-range elastic deformation remain challenging to pursue.
~\footnote{
Here, we wish to remind the reader that the term ``numerically-exact method'' 
stands for a numerical method, in which a model, such as an elastic manifold 
pressed against a corrugated wall with hard-wall repulsion, is solved in such 
a way that all systematic errors can be controlled, e.g., by the number of 
mesh points. 
In contrast, the GW and Persson approaches make uncontrolled approximations.}
This is because treating the wide range of length scales present
in most surface topographies
places great demands on computing time and memory.

Traditionally, many contact mechanics predictions were based on models
following Greenwood and Williamson's (GW) seminal paper~\cite{greenwood66}, 
in which the contact between two solids is treated as the sum of
non-interacting, single-asperity contacts.
The crucial, geometric properties entering such theories 
are the statistics for asperity height and curvature.
One of the most detailed treatments is by Bush et al.~\cite{bush75} who
included a distribution of curvatures and elliptical asperities.
In the last decade,  Persson and collaborators have pursued a 
different 
approach~\cite{persson01jcp,persson02epje,persson02prb2,persson05JPCM}, 
in which
the height autocorrelation function (ACF) is the geometric property
that determines the
pressure distribution and thereby the area of contact.

A central quantity in both GW and Persson's theory
is the ratio of the real area of microscopic contact $A_{\rm c}$ to the
macroscopic projected area of the surfaces $A_0$.
In the limit of low loads, both theories find that
\begin{equation}
\frac{A_{\rm c}}{A_0} = \kappa 
\frac{\sigma_0}{\sqrt{\left\langle \nabla h^2 \right\rangle} E'}
\label{eq:kappa}
\end{equation}
where $\kappa$ is a dimensionless constant of proportionality,
$\sigma_0 \equiv L/A_0$ is the mean pressure normal to the interface,
$\sqrt{\left\langle \nabla h^2 \right\rangle}$ is the root mean-square gradient
of the height profile $h$, and the effective elastic
modulus $E'=E/(1-\nu^2)$, where $E$ is the Young's
modulus and $\nu$ the Poisson ratio.
The generalization of GW by Bush et  al.~\cite{bush75} 
predicts $\kappa=\sqrt{2\pi}$, while Persson theory
yields $\kappa=\sqrt{8/\pi}$.
Numerically exact calculations for the relative contact area
$A_{\rm c}/A_0$ for solids with a self-affine surface topography are, 
give or take, half way between the two theoretical
predictions~\cite{hyun04,campana07epl,hyun07ti}.

Persson's theory also provides a prediction for the functional dependence 
of the stress distribution $P(\sigma)$, which can be described as the
sum of a Gaussian of width $\Delta \sigma$
centered at the macroscopic pressure $\sigma_0$ and
a mirror Gaussian centered at $-\sigma_0$: 
\begin{equation}
P(\sigma) = \frac{1}{\sqrt{2\pi\Delta \sigma^2}}
\left\{ \exp \left[-\frac{(\sigma-\sigma_0)^2}{2\Delta \sigma^2}\right]
- \exp\left[-\frac{(\sigma+\sigma_0)^2}{2\Delta \sigma^2}\right]
\right\}
\label{eq:press_dist}.
\end{equation}
A $\delta$-function contribution is added to $P(\sigma)$ so that
the integral over $P(\sigma)$ from $\sigma=0$ to $\sigma=\infty$ yields unity.
The prefactor to this $\delta$-function contribution 
(divided by the normal macroscopic stress) can be interpreted
as the relative area of the projected surface that is 
{\it not} in contact with the counterface.
For the relevant non-negative pressures, the superposition leads to $P(\sigma)\propto \sigma$ for small $\sigma$ and Gaussian tails at large $\sigma$.
GW theory gives distributions with similar limiting behavior.
Numerical solutions for $P(\sigma)$ are qualitatively consistent with Eq.~\ref{eq:press_dist} when the parabolic peaks of asperities are resolved 
\cite{campana07epl,hyun07ti},
 but some quantitative discrepancies remain as illustrated in the next section.

GW-type theories and Persson theory give quite distinct predictions for the spatial distribution of contact and pressure.
Contact in GW is based on overlap of undeformed surfaces.
This allows one to relate the contact ACF, $C_{\rm c}({\Delta r})$,
to the surface topography.
$C_{\rm c}(\Delta r)$ gives the probability of finding contact at a coordinate 
${\bf r}$ if there is contact at a coordinate ${\bf r}'={\bf r} + \Delta {\bf r}$
that is located at a distance ${\Delta r}=\vert \Delta {\bf r}\vert$ away.
In contrast, Persson's theory contains a prediction for the 
stress ACF, $C_\sigma(\Delta r)$.
Both GW and Persson theory predict power law scaling for the correlation functions, but with very different exponents.
In a numerically exact calculation, Hyun and Robbins found 
that $C_{\rm c}$ and $C_\sigma$ appear to 
decrease algebraically with $\Delta r$ 
in such a way that the two functions are essentially 
proportional to one another~\cite{hyun07ti}.
The values for the exponents that describe the decay of the 
autocorrelation functions were again found to be, give or take, 
half way between the theories.
However, due to computational limitations, the uncertainty on the exponents
was relatively large and only $H=0.5$ was studied.

Here, we would like to reassess these  correlation functions with  Green's
function molecular dynamics (GFMD)~\cite{campana06}, which allows
one to address larger system sizes than with finite-element methods
or multiscale approaches~\cite{yang06epje}.
One of our main goals in these calculations is to assess whether the 
relatively precise values of $\kappa$ predicted by GW and Persson theory 
are to a certain degree fortuitous or if the theories also
predict other computable observables,
specifically stress and contact ACFs,
with similar accuracy, i.e. of order twenty percent. 
We derive approximate and exact bounds on the ratio of these ACFs
and their integrals that supplement the numerical results.
The work presented here
also includes a test of the claim that 
corrections to Persson's theory can be derived with the help
of a systematic expansion scheme derived recently by one of 
us~\cite{muser08prlT}.
The expansion has so far only been worked out (to harmonic order)
for walls that repel each other with forces that increase exponentially 
when the distance between the surfaces decreases. 
Therefore, we also include comparison to numerically exact simulations
based on exponentially repulsive walls.
Finally, we conduct an analysis of connected contact patches
for hard wall interactions
to see if these regions show the characteristic behavior of Hertzian
contact mechanics, as one would expect according
to an overlap theory of purely repulsive, corrugated walls.

The remainder of this paper is organized as follows.
In section~\ref{sec:method}, the GFMD is quickly introduced.
In section~\ref{sec:results}, bounds on the correlations are derived and numerical results for scaling behavior presented.
Conclusions will be drawn in  section~\ref{sec:conclusions}.

\section{Method}
\label{sec:method}

In this work, we model elastic, frictionless
contact between two solids with self-affine surfaces.
Use is made of the mapping of such a system onto
contact between a flat elastically-deformable solid and
a rigid, corrugated substrate~\cite{johnson96}.
The Green's function molecular dynamics (GFMD)
method~\cite{campana06} was used to solve for the elastic response
of the deformable solid.
Details of this method are provided in Ref. \cite{campana06} and we only give
the parameters of the model here.
We chose both Lam{\'e} constants to be unity, which is equivalent to
a Young's modulus of $E = 5/2$, a bulk modulus of $K  = 5/3$, and
a Poisson ratio of $\nu = 1/4$.
In what follows, most stated quantities will be dimensionless,
but we take the Lam{\'e} constants as our unit of pressure.
The continuum equations have no intrinsic length scale, so we will
normalize lengths by the lateral resolution $a$ of the GFMD.

The surface topography of the rigid substrate was a self-affine fractal.
Surfaces with the desired value of the Hurst roughness exponent
$H$ were created using a Gaussian filter technique
for the Fourier components of the height profile
$\tilde{h}({\bf q})$~\cite{meakin}. 
The long wavelength cutoff of fractal scaling $\lambda_{\rm l}$
is always identical to the length of our periodic simulation cell in
both lateral directions.
The effect of reducing $\lambda_{\rm l}$ is discussed in Ref. \cite{hyun07ti}.
Unless noted $\lambda_{\rm l}/a = 4096$.  The effective depth of the deformable solid is also equal to $\lambda_{\rm l}$.
The short wavelength cutoff $\lambda_{s}$ was varied from $a$ to $64 a$.
The wavevectors associated with the cutoffs and resolution are denoted
as $q_{\rm l}=2\pi/\lambda_{\rm l}$,  $q_{\rm s}=2\pi/\lambda_{\rm s}$,
and $q_{\rm a} =2\pi/a$.
The magnitude of the Fourier components was adjusted to maintain
the same mean-square height gradient
$\sqrt{ \left\langle \nabla h^2 \right\rangle} = 0.031$ for all $\lambda_{\rm s}$ and $H$.

\begin{figure}[hbtp]
\begin{center}
\includegraphics[width=10.0cm]{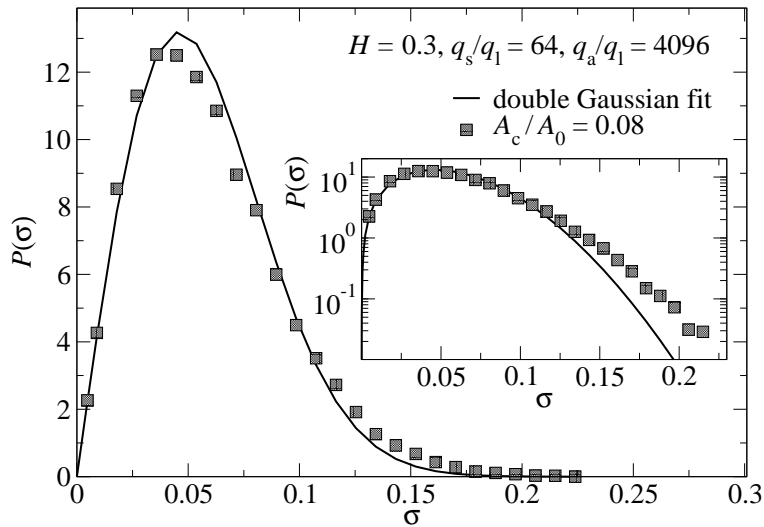}
\end{center}
\caption{\label{fig:press_dis}
Pressure distribution $P(\sigma)$ for a system with $\lambda_{\rm l}=64 \lambda_{\rm s}=4096a$ at a relative contact area of $f_{\rm c} = A_{\rm c}/A_0 = 0.08$.
The line is a fit with Eq.~\ref{eq:press_dist}.
From the area and load we find $\kappa =2.06$ (Eq. \ref{eq:kappa})
with an absolute uncertainty due to finite resolution of less than 0.1.
}
\end{figure}

A wide range of parameters was considered.
The roughness exponent was varied from 0.3 to 0.8, which covers the range of values typically reported in experiments \cite{bonamy06,bouchaud97,dieterich96,krim95}.
The load was varied between $0.001$ and $0.256$,
resulting in fractional contact areas $f_{\rm c} \equiv A_{\rm c}/A_{0}$ between $0.02$ and $0.96$.
For each case we evaluated the spatial variation of the contact pressure
$\sigma({\bf r})$.
Then the total contact area $A_{\rm c}$, probability distribution $P(\sigma)$ of local pressures, and contact and pressure correlation functions were calculated.
Since $P(\sigma)$ enters the analysis below, we show typical results in Figure \ref{fig:press_dis}, along with the analytic expression of Eq. \ref{eq:press_dist}.
For this case $\lambda_{\rm s} >> a$, implying that the top of each asperity is resolved into a smooth parabolic peak.
As a result, $P(\sigma)$ drops as $\sigma$ decreases to zero.
Extending the  roughness to $\lambda_{\rm}=a$ changes the form of $P(\sigma)$
at large and small $\sigma$, but does not change the power law behavior of
the correlation functions at small $q$ that are our focus here \cite{hyun07ti}.

The contact ACF, $C_{\rm c}(\Delta r)$, is defined as
\begin{equation}
C_{\rm c}(\Delta r) = \left\langle
\Theta\{\sigma({\bf r})\} \Theta\{ \sigma({\bf r}')\}
\right\rangle
\end{equation}
where $\Delta r = \vert {\bf r} - {\bf r}' \vert$
and $\sigma({\bf r})$ is the normal component
of the stress at position ${\bf r}$.
\footnote{Note that $\sigma$ is calculated from the force from the rigid substrate normalized by the area, $a^2$, per "atom" in the GFMD.
Our calculation actually uses the $z$ component of this force which differs from the normal component by a factor of the cosine of the angle between the surface normal and the z-axis.  Since the rms slope is only 0.031, the difference is negligible.}
The Heaviside step function, $\Theta(...)$, is zero for negative and unity for positive arguments.
However, unlike the usual convention, we choose $\Theta(0)=0$,
i.e., the step function is unity only 
at those locations where there is contact.
The stress ACF is similarly defined as:
\begin{equation}
C_\sigma(\Delta r) = \left\langle
\sigma({\bf r}) \sigma({\bf r}') \right\rangle,
\end{equation}
The ACF are most conveniently calculated by Fourier transforming.
We
choose the normalization so that 
\begin{equation}
\tilde{C}_\sigma(q) = \left\langle
\tilde{\sigma}^*({q}) \tilde{\sigma}(q) \right\rangle.
\end{equation}

\section{Results}
\label{sec:results}

\subsection{Relation between contact and stress autocorrelation functions}

Traditional models ignore correlations in surface displacement due to elastic deformation so that the location of contacting regions is determined solely by the local height.
We will refer to such models generically as "overlap models".
A particular overlap model is
the bearing area model \cite{johnson96} in which contact occurs wherever the undeformed solids would overlap.
For the case of rough on flat considered in our calculations, this corresponds to the region where the height of the rough solid is above a threshold value.
In the GW model and extensions \cite{greenwood66,bush75}, the same criterion is used to determine which asperities are in contact, and the corresponding load is obtained from the force needed to remove overlap.
Since contact only depends on the local height in such models, 
it is relatively easy to construct the contact morphology
for any given surface topography.
One can then calculate the relative contact area and
$C_{\rm c}(\Delta r)$.

Since the location of contacts is entirely determined by the height in the bearing area and GW model,
Hyun and Robbins argued that $C_{\rm c}$ should have the same scaling as the height-height correlation $C_{\rm h}$ \cite{hyun07ti,meakin}:
\begin{equation}
\tilde{C}_{\rm c} (q) \sim \tilde{C}_{\rm h}(q) \sim q^{-2(1+H)} .
\label{eq:rigid}
\end{equation}
Their numerical results for $H=0.5$ were consistent with this relation.
We are not aware of any specific predictions for the stress ACF in the GW model, although it might also follow $C_{\rm h}$ since the stress on asperities is also determined directly by their height.

In contrast, Persson's scaling theory does not consider $C_{\rm c}$,
but contains implicit predictions for the stress ACF.
In the limit of complete contact, Persson theory becomes exact.
The stress correlation function
can be determined from the height correlation function and
the elastic Greens function.
The latter scales as $q^2$, yielding $\tilde{C}_{\sigma} \sim q^{-2H}$
\cite{greenwood96}.
This result is consistent with numerical studies of full contact \cite{roux93}.
The original version of Persson theory does not discuss stress ACFs
in partial contact explicitly.
One may thus be tempted to use the full contact approximation for the stress
ACF at all loads.
However, when contact is not complete, only a fraction of the elastic
manifold conforms to the counterface and so the full contact ACF only
provides an upper bound, as the non-conforming parts of the manifold
do not carry any load.
After receiving  a preprint of our work,
Persson informed us that a correction factor needs to be included into the
calculation to capture this effect~\cite{perssonnote}.
This correction factor is the relative contact area at a given ``magnification'',
i.e., one needs to correct with the relative contact area $f(q)$ that
his theory would
predict if all roughness for wavevectors of magnitude greater than $q$ were
eliminated.
(This correction factor had already been introduced for the calculation of adhesive interactions in earlier work\cite{persson02epje}.)
With this correction factor the exponent $\mu_{\sigma}=1+H$ is obtained in Persson theory for partial contact.
In particular,
\begin{equation}
\tilde{C}_{\sigma} (q) = 
\underbrace{\frac{E'^2}{4}\,q^2 \tilde{C}_{\rm h}(q)}_{\sim q^{-2H}}
\, \cdot \underbrace{ f(q) }_{\sim q^{-(1-H)}}
\sim  q^{-(1+H)}	
\label{eq:c_s_q}
\end{equation}
where the first term is the full contact prediction and
$f(q) \sim q^{-(1-H)}$
~\cite{persson02epje,perssonnote}.

While the power laws in Equations \ref{eq:rigid} and \ref{eq:c_s_q} are very different, it is not clear how the scaling of the stress ACF and contact ACF should be related.
In the following we argue that two approximate bounds on the ACF's force them to have the same scaling exponents.
In particular,
\begin{equation}
C_{\rm c}(\Delta r) \le \frac{1}{\sigma_{\rm c}^2} \, C_{\sigma}(\Delta r) 
\le \frac{\left\langle \sigma^2 \right\rangle_{\rm c}}{\sigma_{\rm c}^2}  \,
C_{\rm c}(\Delta r) \approx 2C_{\rm c}(\Delta r) ,
\label{eq:inequals}
\end{equation}
where $\sigma_{\rm c} =\langle \sigma \rangle_{\rm c}$ and $\left\langle \sigma^2\right\rangle_{\rm c}$ are the mean and second moments of the
stress averaged over those areas where there is contact, i.e.,
\begin{equation}
\left\langle \sigma^n\right\rangle_{\rm c} = 
\frac{\int_{0^+}^\infty d\sigma \, \sigma^n P(\sigma)}{\int_{0^+}^\infty d\sigma \, P(\sigma)}.
\end{equation}
In cases where the stress histogram,
$P(\sigma) = \langle \delta\{\sigma-\sigma({\bf r})\} \rangle$ can be described
by equation~(\ref{eq:press_dist}), one can easily  
find that $\langle \sigma^2\rangle_{\rm c} /\sigma_{\rm c}^2 =2$.
The same ratio is obtained for the approximately exponential distribution
of pressures found for surfaces with $\lambda_{s}=a$ \cite{hyun04}.
Similar ratios are obtained for the other $\lambda_{\rm s}$ considered here, leading us to add the approximate equality at the right of Eq. \ref{eq:inequals}.

In the limit $\Delta r \to 0$, the stress ACF is exactly equal to
the upper limit of Eq. \ref{eq:inequals}: $C_{\sigma}(0) \equiv \langle\sigma^2\rangle_{\rm c}$.
In the large $\Delta r$ limit, the local values of the stress should become decorrelated, and $C_{\sigma}$ will then equal the lower limit.
One expects a smooth crossover between the two bounds as $\Delta r$ increases
unless there is a strong correlation at some given wavelength.
For example, one can construct counterexamples to the bounds such as a perfectly sinusoidal surface topography.

\begin{figure}[hbtp]
\begin{center}
\includegraphics[width=10.0cm]{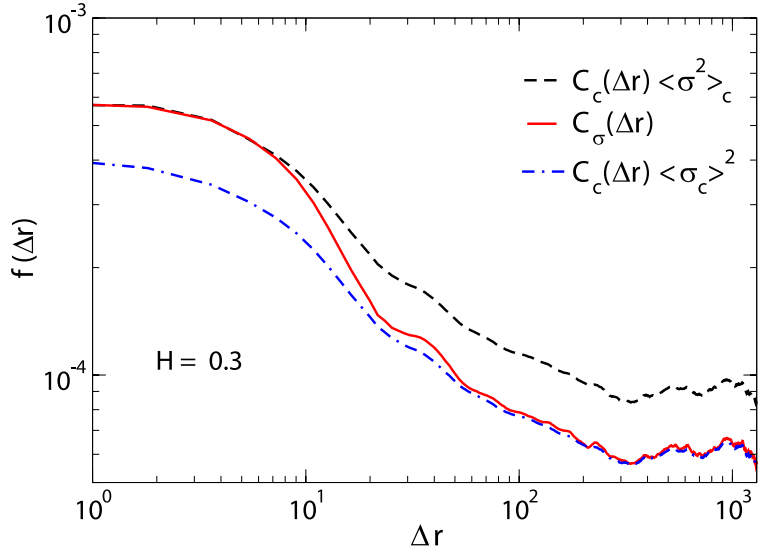}
\end{center}
\caption{\label{fig:C_cc_ss.comp} (Color online)
Stress autocorrelation function (ACF)
$C_\sigma(\Delta r)$ in real space for $H=0.3$ and comparison to the 
upper and lower bound estimates of the stress ACF via 
equation~(\protect{\ref{eq:inequals}}) and the contact ACF $C_{\rm c}$.
Here $\lambda_{\rm l}=64 \lambda_{\rm s}=4096 a$ and $A_{\rm c}/A_0 =0.14$.
}
\end{figure}

Figure~\ref{fig:C_cc_ss.comp} presents typical numerical results for the ACF's of self-affine surfaces.
Values of $C_\sigma$ lie close to the upper bound up to $\Delta r \sim 8a$,
and then cross over rapidly to the lower bound.
A heuristic reason for the more rapid drop in $C_\sigma$ than $C_{\rm c}$ can be constructed as  follows.
Consider the contribution to these two ACFs that stem from two points ${\bf r}$ and ${\bf r}'$ that are in the same contact patch, for example, within a single Hertzian contact.
The contribution to $C_{\rm c}(\Delta r)$ will simply
be unity, i.e., $C_{\rm c}(\Delta r)$ cannot decay
within a simply-connected contact region.
Conversely,
$C_\sigma(\Delta r)$ can and will have a lot of structure, e.g.,
the correlation between the stress at the center and edge of a Hertzian contact will be small.
Consequently, a significant fraction of $C_\sigma(\Delta r)$
will have decayed  on a length-scale $\Delta r$ that is comparable
to a typical contact radius, while only a very small fraction of
$C_{\rm c}(\Delta r)$ will have decayed on that same distance.
Note that the rapid decay in Fig. \ref{fig:C_cc_ss.comp} starts at
$\Delta r/\lambda_{\rm s} \sim 1/8$ which should be comparable to the smallest contacts.
Since contacts of many different sizes are found, one may conjecture that $C_\sigma(\Delta r)$ falls off faster than $C_{\rm c}(\Delta r)$ for all $\Delta r$.
The difference should decrease at large $\Delta r$ as the number of connected clusters with dimension greater than $\Delta r$ decreases.

\begin{figure}[hbtp]
\begin{center}
\includegraphics[width=10.0cm]{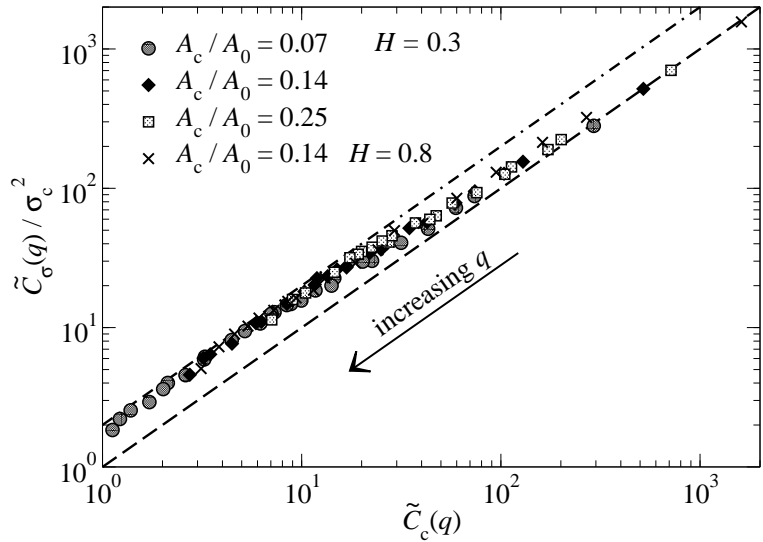}
\end{center}
\caption{\label{fig:C_ss_Cqq} The stress autocorrelation function
$\tilde{C}_\sigma(q)$
normalized by $\sigma_{\rm c}^2$ as a function of
$\tilde{C}_{\rm c}(q)$.
The two straight lines reflect the inequalities from 
equation~(\protect{\ref{eq:inequals}}).
}
\end{figure}

Fig~\ref{fig:C_ss_Cqq} illustrates that the same upper and lower bounds describe the Fourier transforms of the correlation functions.
Data for a range of $H$ and relative areas are shown.
In each case there is a crossover from the lower bound of Eq. \ref{eq:inequals} at small $q$ (large $\tilde{C}$) to the upper bound at large $q$.

One can derive additional relations for the Fourier transforms of the ACF's that support the bounds quoted above.
First, the $q=0$ values must obey:
\begin {eqnarray}
\tilde{C}_{\rm c}(q=0)&=&|\Sigma_{\bf r} \Theta(\sigma({\bf r}))|^2 = \left[ N A_{\rm c}/A_{0}\right ]^2 \\
\tilde{C}_{\sigma}(q=0)&=&|\Sigma_{\bf r} \sigma({\bf r})|^2 = \left[N \sigma_{\rm c} A_{\rm c}/A_{0} \right]^2  \ \ ,
\label{eq:qequal0}
\end{eqnarray}
where $N$ is the number of $\bf r$ 
(or real-space grid points)
in the sum.
This shows that in the limit $q \rightarrow 0$ the correlation functions satisfy the lower bound in Eq. \ref{eq:inequals}.
One can also use a general sum rule over real and reciprocal space to write:
\begin{equation}
\Sigma_q \tilde{C}_{\sigma}(q)=\Sigma_{q} |\sigma(q)|^2 = N \Sigma_{\bf r} |\sigma({\bf r})|^2 = \langle \sigma^2 \rangle_{\rm c} 
 N^2 A_{\rm c}/A_{0}
\label{eq:sumcs}
\end{equation}
and similarly
\begin{equation}
\Sigma_q \tilde{C}_{\rm c}(q)= N^2 A_{\rm c}/A_{0} .
\label{eq:sumcc}
\end{equation}
The last two equations show that the sum over all $q$ of the ACFs is exactly consistent with the upper bound in Eq. \ref{eq:inequals}.
While this implies that the upper bound must be exceeded for some $q$, we will see that the sum is dominated by large $q$ where the upper bound is nearly obeyed.
Our focus is on the power law scaling regime in the opposite limit of small $q$.

In order to collapse data for different loads it is useful to recast the above sum rules.
It is also helpful to eliminate the $q=0$ term since we will see that the ACFs diverge in the limit $q \rightarrow 0$.
In addition, the $q=0$ term scales as $(A_c/A_0)^2$, while other terms
scale linearly with $A_c/A_0$.
Subtracting Eq. \ref{eq:qequal0} from Eq. \ref{eq:sumcc} and rearranging factors we find:
\begin{equation}
\Sigma_{q\neq 0} f_{\rm c} \tilde{C}_{\rm c}(q)/(1-f_{\rm c})\tilde{C}_{\rm c}(0) =1 ,
\label{eq:normal}
\end{equation}
where $f_{\rm c} = A_{\rm c}/A_{0}$.
We will see that scaling $\tilde{C}_{\rm c}$ in this way removes the dependence on $f_{\rm c}$ at small $f_{\rm c}$.
Evaluating the same weighted sum for the stress ACF yields
\begin{equation}
\Sigma_{q\neq 0} f_{\rm c} \tilde{C}_{\sigma}(q)/(1-f_{\rm c})\tilde{C}_{\sigma}(0) =
[\langle \sigma^2 \rangle_{\rm c} 
 / \sigma_{\rm c}^2 -f_{\rm c}]/[1-f_{\rm c}] >1.
\end{equation}
Using the same scaling of the two ACF's guarantees that they coincide as $q \rightarrow 0$.
While $\tilde{C}_{\sigma}$ will decay more slowly at large $q$,
the fact that its integral over all $q$ is larger by only a factor of order 2
for any system size implies that $\tilde{C}_{\sigma}$ should be described by the same scaling exponent as $\tilde{C}_{\rm c}$.

\subsection{Comparison to overlap theories}

\begin{figure}[hbtp]
\begin{center}
\includegraphics[width=10.0cm]{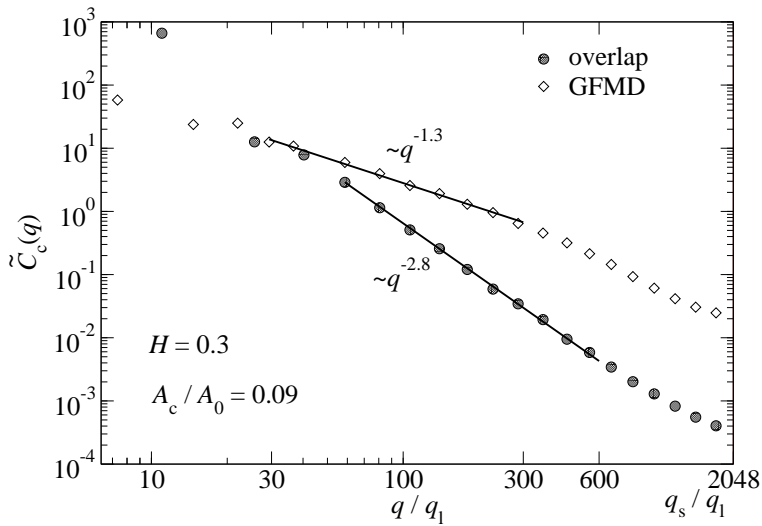}
\end{center}
\caption{\label{fig:sigma_c_GW_H0_3} 
The contact ACF, $\tilde{C}_{\rm c}(q)$,
as a function of wavevector $q$ divided
by the long wavelength cutoff $q_{\rm l}$.
Predictions from the bearing area model and GFMD calculations are shown
for a Hurst roughness exponent of $H=0.3$.
Here $\lambda_{\rm l}=2048 \lambda_{\rm s}=2048 a$
and $A_{\rm c}/A_0=0.09$.
The lines reflect power laws in $q$.
}
\end{figure}

Figure~\ref{fig:sigma_c_GW_H0_3} compares results for $\tilde{C}_{\rm c}$ from the full GFMD calculation and the GW model which uses overlap to determine contact.
Both models yield roughly power law behavior at intermediate $q$
\begin{equation}
\tilde{C}_{\rm c}(q) \propto q^{-\mu_{\rm c}}  .
\end{equation}
However there is a large discrepancy in the numerical values
between GFMD ($\mu_{\rm c}\approx 1.3$) and the bearing area model 
($\mu_{\rm c} \approx 2.8$), which one can conclude from 
figure~\ref{fig:sigma_c_GW_H0_3}.
A similar difference in the values for $\mu_{\rm c}$ is observed for
other $H$ and $\lambda_{\rm s}$, as reported for $H=0.5$ in Ref.~\cite{hyun07ti}.
There is a simple physical reason that overlap models yield larger $\mu_{\rm c}$ and thus cluster the contact patches too closely.
They neglect the fact that an asperity in the
vicinity of a very high asperity is less likely to
come into contact because it is pushed down when that very
high asperity comes in contact with the 
counter-surface~\cite{persson02prb2,hyun04,hyun07ti,campana07ap}.

In all cases, our numerically determined exponent for the bearing area model is consistent with the estimate $\mu_{\rm c} = 2(1+H)$, within our uncertainty of $\sim 0.25$ for this model.
In contrast, the GFMD results are consistent with $\mu = 1 + H$ within an uncertainty of $\sim 0.1$.
The reason for the difference in the uncertainties of our estimates stems from the fact that contact geometries associated with large $\mu_{\rm c}$ have large finite-size effects, particularly when $\mu_{\rm c}$ exceeds two.
\footnote{
We found $\mu_c$ was 3.1 for $\lambda_{\rm l}/\lambda_{\rm s} = 64$ and 2.8 for $\lambda_{\rm l}/\lambda_{\rm s} = 2048$ for the bearing area model, while the size effect was merely 0.1 for the GFMD data.}
This is because a large value of $\mu_{\rm c}$ implies significant contributions from long wavelengths, which have less sampling than short wavelengths, as can be seen from an analysis of $C_{\rm c}({\bf r})$:
\begin{eqnarray}
C_{\rm c}({\bf r})& =& \int d^2 q \, \tilde{C}_{\rm c}({\bf q}) \exp(-i{\bf q}\cdot {\bf r}) \\
& \sim & \int_a^{\lambda_{\rm l}} dq q^{1-\mu_{\rm c}} \int_0^{2\pi} d\theta \exp(-iqr \cos (\theta))  \ \ .
\label{eq:convert}
\end{eqnarray}
For $\mu_{\rm c} < 2$ the contribution at small $q$ is finite.
One can change the integration variable to $qr$ and finds:
\begin{equation}
C_{\rm c}(\Delta r)-C_{\rm c} (\infty) \propto \Delta r^{-\nu_{\rm c}} 
\end{equation}
at large $\Delta r$ with $\nu_{\rm c} = 2-\mu_{\rm c}$.
This is the type of behavior we find for our numerical solutions,
i.e. Fig. \ref{fig:C_cc_ss.comp}.

The behavior becomes qualitatively different if $\mu_{\rm c}>2$ as predicted by the
$\mu_{\rm c} =2(1+H) >2$ relation for the bearing area model.
This implies a singular contribution from small $q$ in Eq.~\ref{eq:convert} and suggests that $C_{\rm c} (\Delta r)$ should increase with $\Delta r$.
The origin of this behavior seems to be related to the distribution of sizes of connected regions in the bearing area model.
The probability that a cluster will have area $a_{\rm c}$ is predicted~\cite{meakin} to scale as $P(a_{\rm c}) \sim  a_{\rm c}^{-\tau}$ with $\tau=(2-H/2)$.
Since $\tau <2$, most of the contact area is in the clusters of largest size, and there will be no decay in $C_{\rm c} (\Delta r)$ on this scale.
The dominance of large clusters leads to significant fluctuations in data for the bearing area model that are not seen in the numerical solution with GFMD.
We conclude that the bearing area model and thus GW provide a very poor description of the contact ACF.

\subsection{Numerical Determination of $\mu$}

\begin{figure}[hbtp]
\begin{center}
\includegraphics[width=10.0cm]{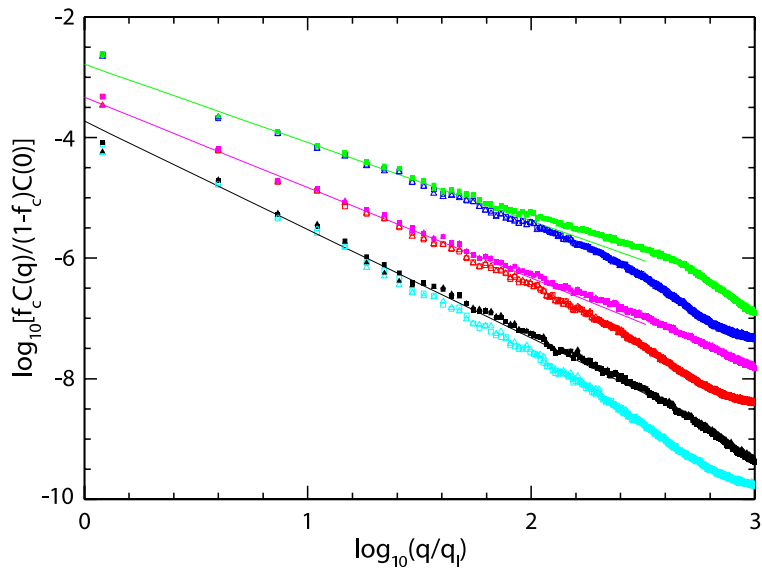}
\end{center}
\caption{\label{fig:powerlaws} 
(Color online)
Comparison of the stress (open symbols) and contact (closed symbols) ACF's for $H=0.3$, 0.5 and 0.8 from top to bottom.
In each case results for $f_{\rm c}$ near 4\% (triangles) and 8\% (squares) are collapsed by plotting $f_{\rm c} \tilde{C}(q)/(1-f_{\rm c})\tilde {C}(0)$.
Curves for $H=0.5$ and $0.8$ are offset vertically downward by successive factors of ten to prevent overlap.
Straight lines are fits with $\mu = 1+H$.
In all cases $\lambda_{\rm l}=2048 a$ and $\lambda_{\rm s}=a$.
}
\end{figure}

To determine accurate values of the scaling exponents describing the ACF's we maximized the scaling region by taking $\lambda_{\rm s}=a$.
Our results and earlier work \cite{hyun07ti} show that resolving the asperities is not important to the large scale behavior of interest here.
Figure \ref{fig:powerlaws} shows results for $\tilde{C}_{\sigma}$ and
$\tilde{C}_{\rm c}$ at $H=0.3$, $H=0.5$ and $H=0.8$.
In each case, data for two different area fractions, corresponding to about 4\% and 8\% have been collapsed by plotting $f_{\rm c} \tilde{C}(q)/ (1-f_{\rm c}) \tilde{C}(0)$ following Eq. \ref{eq:normal}.
With this rescaling, the data for the two loads are indistinguishable.
This is consistent with previous numerical work that indicates area fractions of less than 10\% exhibit scaling behavior consistent with the asymptotic small $f_{\rm c}$ limit \cite{hyun04,hyun07ti,campana07ap}.

At small $q$ the stress and contact ACF follow the same scaling.
Straight lines show that the data are consistent with $\mu_{\rm c}=\mu_{\sigma} = 1+H$ in this regime.
This ansatz for the scaling exponent was motivated by the fact that $\mu$ is bigger than the value $2H$ predicted for full contact
by an amount that decreases with increasing $H$ (see below).
It also ensures that $\mu$ remains below 2 as $H$ increases to unity,
and thus that the real space correlation function has nonsingular scaling behavior.
If both stress and contact ACF are assumed to have the same exponent in numerical fits to the data, then the deviation from $1+H$ is less than 0.1
(Table \ref{tab:exponents}).
However, the fact that $\tilde{C}_{\rm c}$ always decays by about an extra factor of two means that separate fits always give $\mu_{\rm c} > \mu_{\sigma}$.
Given that our scaling range is only a decade and a half, the magnitude of this difference is of order $\log_{10}(2)/1.5 \approx 0.2$.
We cannot rule out deviations between $\mu_{\sigma}$ and $\mu_{\rm c}$ on this scale and it represents the major source of uncertainty in Table \ref{tab:exponents}.
Given this uncertainty, it is possible that $\mu_{\rm c}$ may reach or exceed 2 before $H$ reaches unity.
As for overlap models, this would imply singular contributions from large contacts, and anomalous behavior of the correlation function at large $\Delta r$.

\begin{table}[hbtp]
\begin{center}
\begin{tabular}{|l|l|l|l|}       \hline
        & GFMD & GW & Persson \\ \hline
$H=0.3$ &  $\mu = 1.28(7)$ & $\mu_{\rm c} = 2.6$   & $\mu_{\sigma} = 1.3$   \\ \hline
$H=0.5$ &  $\mu = 1.52(7)$ & $\mu_{\rm c} = 3.0$   & $\mu_{\sigma} = 1.5$   \\ \hline
$H=0.8$ &  $\mu = 1.86(12)$  & $\mu_{\rm c} = 3.6$  &  $\mu_{\sigma} = 1.8$   \\ \hline
\end{tabular}
\end{center}
\caption{\label{tab:exponents}
Summary of the exponents found for the stress/contact ACFs
at small $f_c$ for different roughness exponents $H$
and the different methods
analyzed in this work.  Results for GW and Persson \cite{perssonnote}
are analytic predictions.
GFMD results are numerical fits assuming $\mu_{\sigma}=\mu_{\rm c}$ and the numbers in parentheses are uncertainties in the last significant digit.}
\end{table}

\subsection{Comparison to Persson theory}

\begin{figure}[hbtp]
\begin{center}
\includegraphics[width=10.0cm]{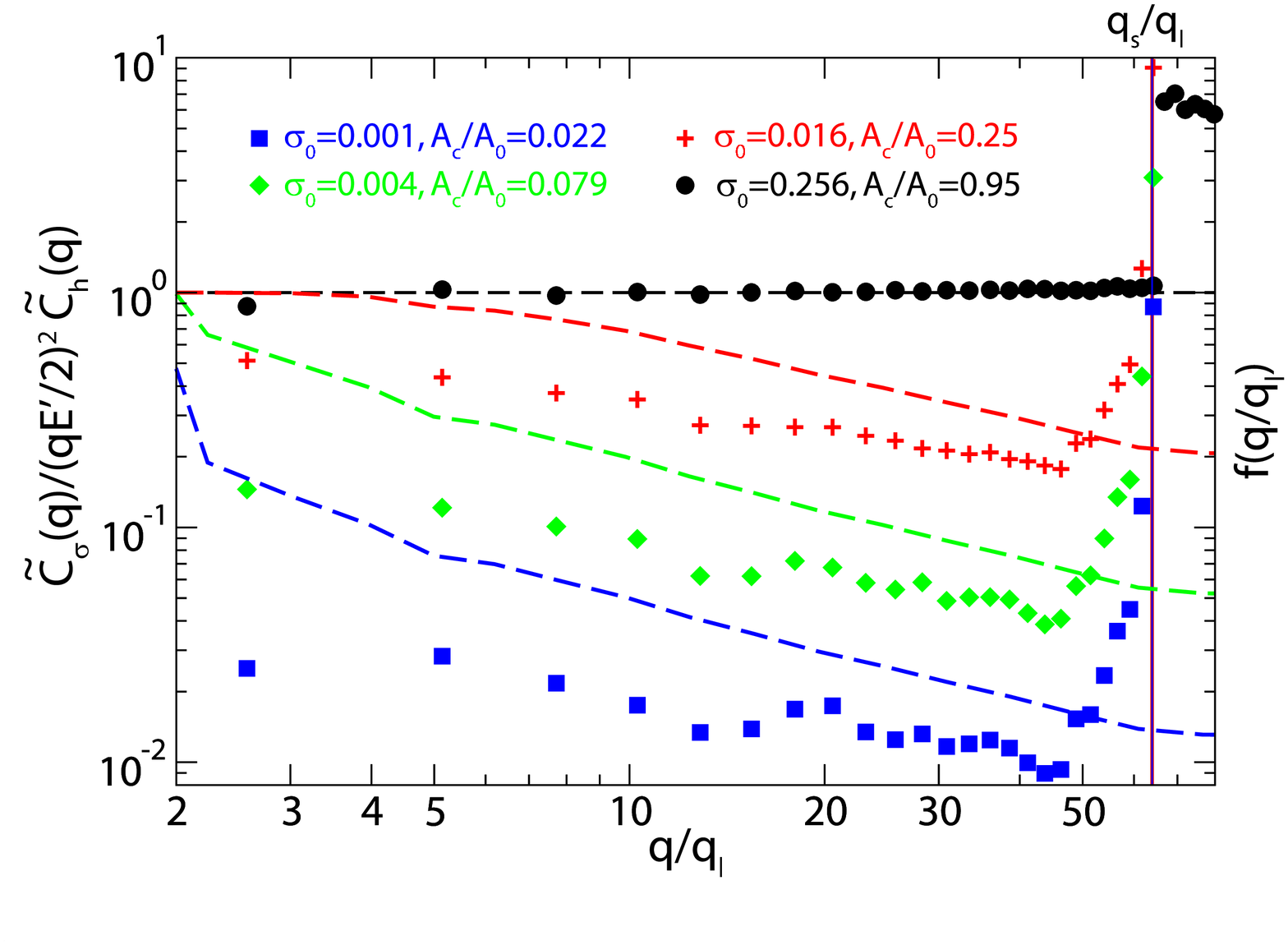}
\end{center}
\caption{\label{fig:comp_Pers_H0_3} 
(Color online)
The stress ACF,
$\tilde{C}_\sigma(q)$,
normalized by the stress ACF for full contact
as a function of $q/q_{\rm l}$
for different relative contact areas $A_{\rm c}/A_0$ and
roughness exponent $H=0.3$.
Results were averaged over the direction of ${\bf q}$ and over small bins in the magnitude to reduce numerical scatter. 
From Eq. \ref{eq:c_s_q} this ratio should be $f(q)$,
the relative contact area at a given magnification $q/q_{\rm l}$.
Dashed lines show a numerical evaluation of this quantity
for the same surfaces.
}
\end{figure}

\begin{figure}[hbtp]
\begin{center}
\includegraphics[width=10.0cm]{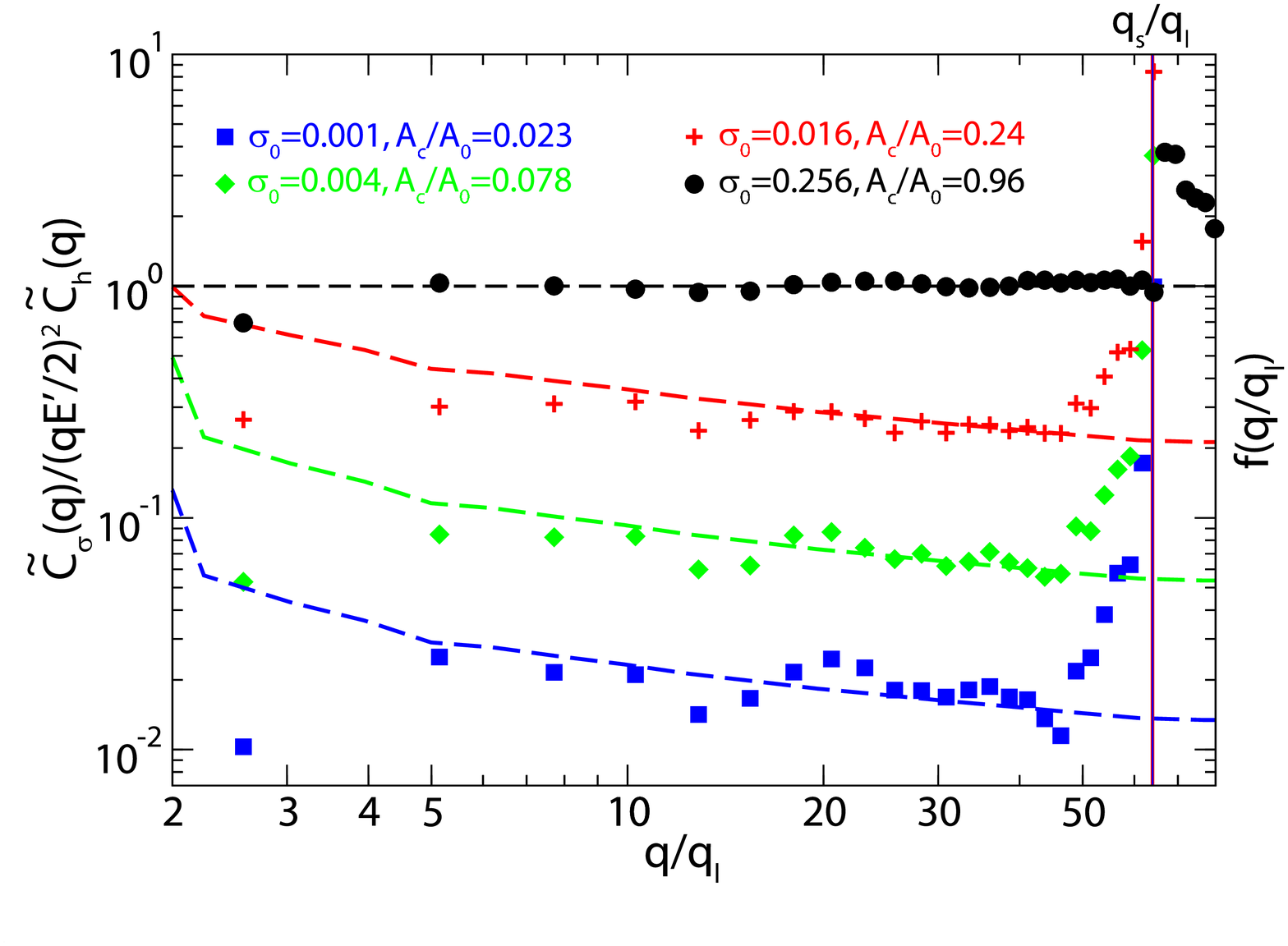}
\end{center}
\caption{\label{fig:comp_Pers_H0_8} 
(Color online.)
Same as previous figure, except that now the roughness exponent
is $H=0.8$ instead of $H=0.3$.
Note that all configurations were produced with the same random seed,
which explains the correlation in the noise for different $A_{\rm c}/A_0$.
}
\end{figure}

When comparing the GFMD results to Persson's theory, it is more
convenient to compare the stress ACFs rather than the contact ACFs.
Figures~\ref{fig:comp_Pers_H0_3} and \ref{fig:comp_Pers_H0_8} show
our numerical results for $\tilde{C}_{\sigma}(q))$ normalized by
the stress ACF from the full contact approximation ($f(q)=1$ in
Eq. \ref{eq:c_s_q}).
Results are shown
for a wide range of loads that give relative contact area
$f_{\rm c}$ between 2\% and 96\%.
Note that theory and simulation should not be compared for $q > q_s$
where $C(q) \rightarrow 0$ and the plotted ratio is not well defined.

Our numerical results for $f_{\rm c} > 0.9$ are nearly indistinguishable
from the full contact expression.
As $f_{\rm c}$ decreases, the discrepancies from full contact increase.
The magnitude of $\tilde{C}_{\sigma}(q)$ is depressed and the
data drop with an apparent power law indicating that $\mu > 2H$.
The deviation is clearly smaller for $H=0.8$ than $H=0.3$, but is present in both cases.

In Persson's theory with corrections for partial contact (Eq. \ref{eq:c_s_q})
the ratio plotted in Figs. \ref{fig:comp_Pers_H0_3} and \ref{fig:comp_Pers_H0_8}
should equal $f(q)$, the relative contact area for surfaces resolved to $q$
~\cite{persson02epje,perssonnote}.
Dashed lines in the figures plot the numerically determined $f(q)$ for
the same surfaces.
For smaller relative contact area, the theory still captures the trend 
correctly, i.e., the power laws with which the ACFs decay at large
$q$ are very similar to the numerical data.
The theoretical prefactors are generally better for large $H$ than for 
small $H$, which is consistent with the observation that the value 
for $\kappa$ predicted by Persson improves with increasing $H$~\cite{hyun04}.

For a few of the largest area fractions
(not included here to make the figures clear),
there appears to be a crossover at a wave vector $q^*$.
For $q< q^*$ the results converge to the full contact prediction,
while for $q >q^*$ the results follow the larger exponent observed at
small area fractions.
It is intriguing to note that
the crossover behavior is observed in the simulations only
near and above the percolation probability for a square lattice $f_{\rm c} \approx 0.59$ \cite{stauffer}.
This suggests that when the contacting area percolates, the system behaves as if it were in full contact at large scales.
The wavelength corresponding to $q^*$ might then correspond to the size of the largest non-contacting regions, leading to an increase in $q^*$ as the area fraction increases.
Tests of these conjectures are beyond the scope of this work.

\subsection{Comparison to field-theoretical approach}

In reference~\cite{muser08prlT}, it was argued by one of us (MHM)
that Persson's theory corresponds to the leading-order term of a 
rigorous field-theoretical expansion.
The expansion is formally based on the assumption that
a (free) energy functional exists describing the interaction between two
contacting solids, which depends on the gap separating the two solids.

For exponential repulsion, corrections to Persson's theory were
worked out explicitly up to harmonic order.
The main result relevant for the calculation of the correlation function
is that equation~(\ref{eq:c_s_q}), which is  
valid in Persson's theory, will be replaced with
\begin{equation}
\langle \tilde{\sigma}^*({\bf q})\tilde{\sigma}({\bf q})\rangle = 
\left\{ \tilde{G}_1({\bf q}) \right\}^2 \, \frac{E'^2}{4}  
q^2 \langle \tilde{h}^*({\bf q}) \tilde{h}({\bf q}) \rangle.
\end{equation}
Here, $\tilde{G}_1({\bf q}) = 1/(1+\zeta q E'/2\sigma_0)$ is a correction factor
that depends on the characteristic screening length $\zeta$ of the
exponential repulsion, the magnitude $q$ of the wavevector, the
effective elastic constant $E'$ and the macroscopic normal stress $\sigma_0$.
As the normal force is never exactly zero, we have used $f(q)=1$ for all
values of $q$ in the evaluation of Eq.~(\ref{eq:c_s_q}).
Thus, like Persson theory, the field theory has no adjustable coefficient.
In the limit $\zeta\to 0$, corrections disappear.

\begin{figure}[hbtp]
\begin{center}
\includegraphics[width=10.0cm]{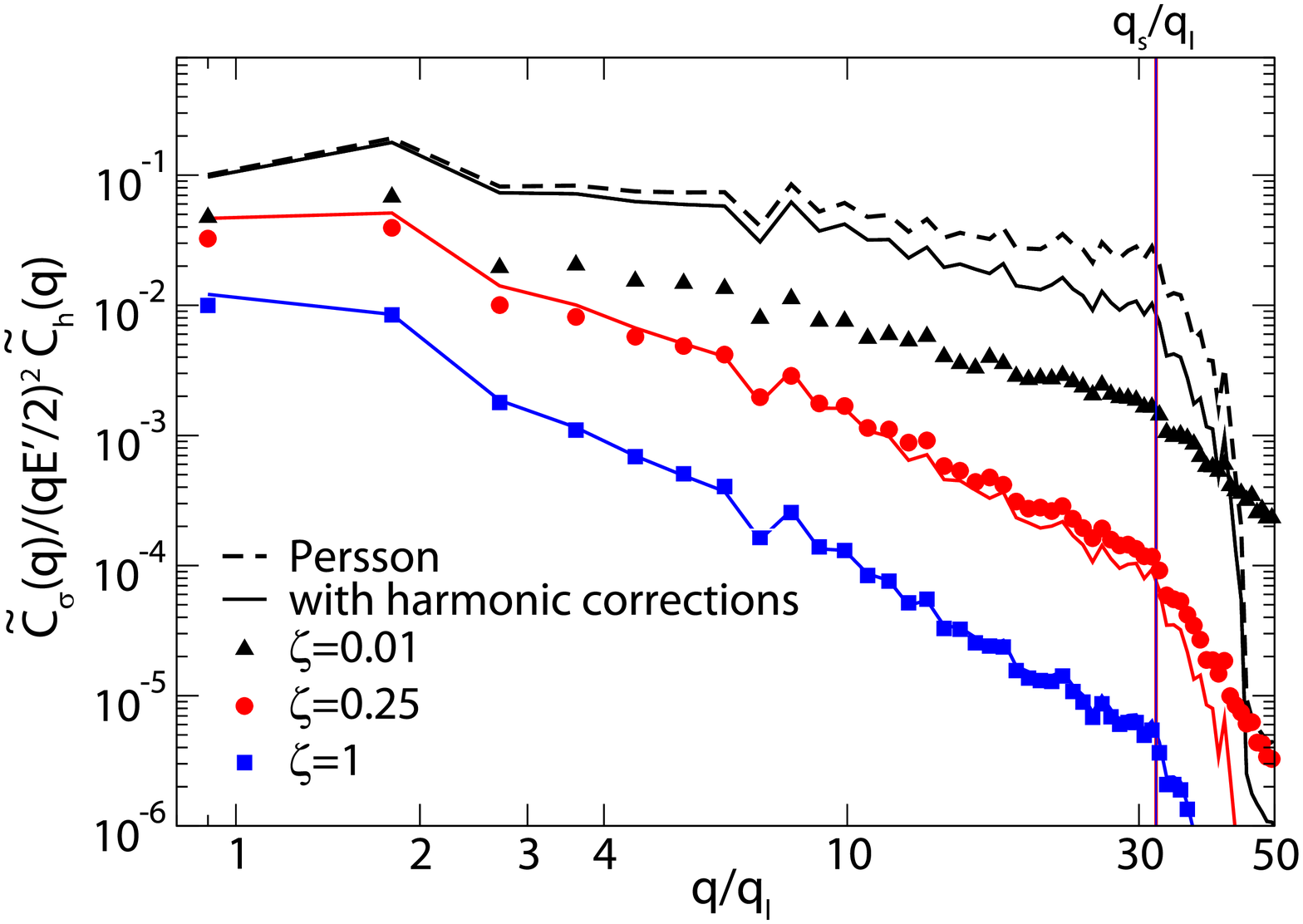}
\end{center}
\caption{\label{fig:u_q_comp_all03}
(Color online)
Calculation of stress ACFs for exponentially repulsive walls.
Symbols indicate simulation data for different values of the
screening length $\zeta$ as measured in the units of the distance $a$
between two discretization points.
The dashed line reflects Persson's theory.
Solid lines are drawn according to the field-theoretical
approach presented in reference~\protect{\cite{muser08prlT}}.
Neither theory has adjustable coefficients.
Here $\lambda_{\rm l}=1024a$, $\lambda_{\rm s} = 32a$, $H=0.3$, and $L/A_0 = 0.004$.
}
\end{figure}

In figure~\ref{fig:u_q_comp_all03} comparison is made between
numerical results and the field theory.
The numerical data was based on the same calculations as
those presented in reference~\cite{muser08prlT}.
It can be seen that corrections to Persson theory are very small
for the smallest value of the screening length analyzed here, 
i.e., for $\zeta=0.01$.
However, for a value of $\zeta=0.25$ the agreement between
predicted and calculated stress ACF is essentially perfect for
the relevant wave vectors.
The degree of agreement is surprisingly good, given the relatively
poor agreement in the stress histograms for that same value,
see figure~1 in reference~\cite{muser08prlT}.
At the largest value of $\zeta$, that is, $\zeta=1$, there is
almost perfect agreement, as to be expected for a rigorous expansion
that is most accurate for large values of $\zeta$.

\subsection{Analysis of connected contact patches}

A merit of GW and related theories is that they provide an intuitive explanation for why total load $L$ and true area of contact $A_{\rm c}$ are proportional to one another at small loads and thus that $\sigma_{\rm c}$ is constant.
Although the relation between area $a_c$ and local load $l_{\rm c}$ for any individual contact is non-linear in these theories, $a_{\rm c} \propto l_c^{2/3}$, the number of contacts of each area rises linearly with load.
As the total load increases, a contact that already exists will have 
an increased local load and grow in size. 
However, new contacts will form under increasing $L$ so that 
there is a supply of new contact patches with small local loads.
Under certain favorable conditions, the distribution of contact loads and sizes maintains the same shape and $\sigma_{\rm c} = L/A_{\rm c}$  remains constant.
Previous work shows that while the distribution of contact areas is different than the GW prediction, it is independent of load at small loads \cite{hyun04}.
Here we examine the relation between load and area within these patches.

\begin{figure}[hbtp]
\begin{center}
\includegraphics[width=10.0cm]{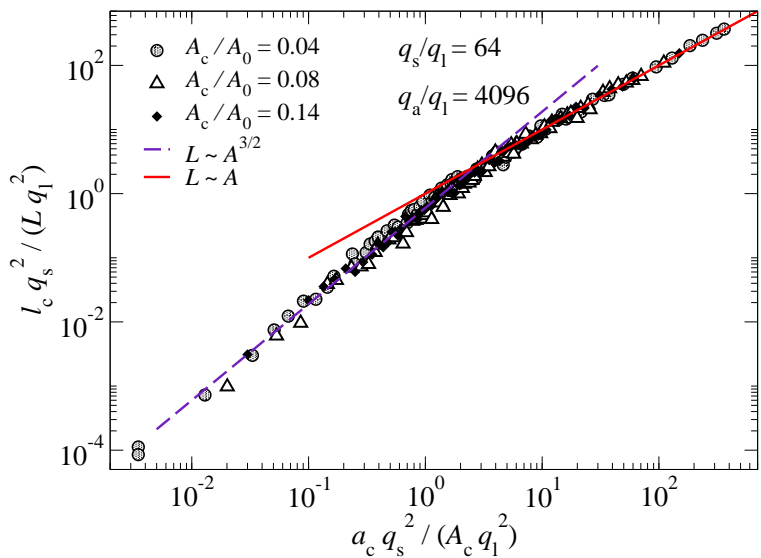}
\end{center}
\caption{\label{fig:L_vs_A_comparisson_4096_collapse_diff_A_real}
Load $l_{\rm c}$ that an individual connected contact patch carries as a
function of its microscopic surface area $a_{\rm c}$.
Here $H=0.3$, $\lambda_{\rm l}=64\lambda_{\rm s}=4096a$ and the area fraction is indicated in the figure.
}
\end{figure}

In figure~\ref{fig:L_vs_A_comparisson_4096_collapse_diff_A_real}
we present data for a number of systems with $H=0.3$.
In particular we analyze
the load $l_{\rm c}$ that connected contact patches carry as a function of their area $a_{\rm c}$.
It can be seen that the data decomposes into two scaling regimes.
At small $a_{\rm c}$, the data follow the prediction of Hertzian contact mechanics that is used in GW, $a_{\rm c} \propto l_{\rm c}^{2/3}$. 
At large $a_{\rm c}$, the load exhibits the linear scaling with area that is found for the entire macroscopic contact area.
The crossover occurs when the contact area is comparable to the square of the small wavelength cutoff in the fractal scaling.
At smaller scales single asperities are fully resolved, while at larger scales one sees the effects of roughness with many wavelengths.
GW theory does not include the effect of these larger scales and it is interesting that the linear scaling of area and load enters so close to $\lambda_{\rm s}$.
Hyun and Robbins have shown that there is also a crossover in the the probability of finding clusters of a given size at $a_{\rm c} \sim \lambda_{\rm s}^2$.
The probability is nearly flat for $a_{\rm c} < \lambda_{\rm c}^2$ and falls off as a power law at larger $a_{\rm c}$ .

\section{Discussion and Conclusions}
\label{sec:conclusions}
In this work, we computed the stress and contact ACFs that one obtains when
pressing an elastically deformable solid against a rough, rigid, 
non-adhesive, and impenetrable substrate.
Analytic arguments were presented for approximate bounds and exact sum
rules on the stress and contact ACFs.
These imply that $\tilde{C}_{\sigma}$ decays more rapidly with wavevector $q$ than $\tilde{C}_{\rm c}$, but that the change in their ratio is only about a factor of two, no matter how large the system.
As a result, the decrease of these ACFs with wavevector
must be described by the same scaling exponent $\mu$.

Numerical results for $\mu$ were compared to the predictions of analytic
theories for roughness exponents $0.3 \leq H \leq 0.8$ that span the typical range for experimental surfaces (Table~\ref{tab:exponents}).
Our GFMD results are consistent with $\mu = 1 + H$ within an error of 0.1.
This is only half the value predicted by the bearing area and GW models, which neglect the elastic interactions between deforming asperities.
Persson's theory includes these correlations through an approximation for the stress ACF that becomes exact in the limit of full contact. 
Recent extensions of his model\cite{persson02epje,perssonnote} that include corrections
for partial contact lead to a scaling exponent that is consistent
with our numerical data \cite{perssonnote}.
The prefactor predicted by this model appears to be slightly too high
at small $H$, but to approach the numerical results as $H \rightarrow 1$.
This is also the limit where the assumption of full contact is most accurate.
In this context it is interesting to note that the value of $\kappa$ 
(Eq. \ref{eq:kappa}) also seemed to approach Persson's prediction 
as $H \rightarrow 1$ in earlier work \cite{hyun04}.

In this work we also tested whether a recently suggested field-theoretical
approach to contact mechanics allows one to improve predictions for the
stress ACFs \cite{muser08prlT}.
The new approach can be interpreted as an expansion, in which the
perturbation parameter is a screening length which describes the
exponential repulsion between two surfaces.
A zero screening length corresponds to hard wall interactions.
In this limit the leading order term of the expansion reduces to Persson's
theory.
Our numerical results suggest that including the next non-vanishing term
vastly improves the agreement between calculated and predicted stress ACFs.

Lastly, analysis of the load that is carried by connected contact patches
revealed a crossover at a critical patch size.
Smaller contacts exhibit a Hertzian relation between area and load.
Larger contacts exhibit a linear relation between area and load.
This linearity at the contact scale may be part of the reason that
a linear relation between area and load is observed for the entire
surface. 

Acknowledgments: MHM thanks Matt Davison for helpful discussions.
Computing time from SHARCNET as well as financial support from NSERC, 
General Motors and National Science Foundation Grant No. DMR 0454947 are gratefully acknowledged.
\\


\end{document}